\documentclass[notitlepage,prd,twocolumn,preprintnumbers,longbibliography,nofootinbib,showpacs]{revtex4-1}
\usepackage[colorlinks=true,urlcolor=blue,linkcolor=blue,citecolor=blue]{hyperref}
\usepackage{slashed,color,amsmath,amssymb,mathrsfs}
\usepackage{amsfonts,epsfig,amssymb}
\usepackage{graphicx}
\usepackage{hyperref}
\usepackage{longtable}
\usepackage{array}
\usepackage{accents}
\usepackage{tensor}
\usepackage{footmisc}
\usepackage{longtable}
\usepackage{booktabs}
\usepackage{multirow}
\usepackage{bm}
\usepackage{subfigure}
\usepackage{gensymb}

\newcommand{\fslash}[1]{{}#1\!\!\!/}

\allowdisplaybreaks

\begin{document}

\title[]{Study of the time-like electromagnetic form factors of the $\Lambda_c$ }
\author{Di Guo$^{1, }$}
\author{Qin-He Yang$^{1,2}$}
\email{yqh@hnu.edu.cn}
\author{Ling-Yun Dai$^{1,2}$}
\email{dailingyun@hnu.edu.cn}
\affiliation{$^{1}$ School of Physics and Electronics, Hunan University, Changsha 410082, China}
\affiliation{$^{2}$ Hunan Provincial Key Laboratory of High-Energy Scale Physics and Applications, Hunan University, Changsha 410082, China}

\date{\today}
\begin{abstract}
In this paper, the reaction of electron-positron annihilation into $\Lambda_c^+\bar{\Lambda}_c^-$ is investigated. The $\Lambda_c^+\bar{\Lambda}_c^-$ scattering amplitudes are obtained by solving the Lippmann-Schwinger equation.
The contact, annihilation, and two pseudoscalar-exchange potentials are taken into account in the spirit of the chiral effective field theory.
The amplitudes of $e^+e^-\to \Lambda_c^+\bar{\Lambda}_c^-$ are constructed by the distorted wave Born approximation method, with the final state interactions of the $\Lambda_c^+\bar{\Lambda}_c^-$ re-scattering implemented. By fitting to the experimental data, the unknown couplings are fixed, and high-quality solutions are obtained.
With these amplitudes, the individual electromagnetic form factors in the timelike region, $G_E^{\Lambda_c}$, $G_M^{\Lambda_c}$, and their ratio, $G_E^{\Lambda_c}/G_M^{\Lambda_c}$, are extracted. Both modulus and phases are predicted. 
These individual electromagnetic form factors reveal new insights into the properties of the $\Lambda_c$.
The separated contributions of the Born term, contact, annihilation, as well as the two pseudoscalar exchange potentials to the electromagnetic form factors are isolated. It is found that the Born term dominates the whole energy region. The contact term plays a crucial role in the enhancement near the threshold, and the annihilation term is essential in generating the fluctuation of the electromagnetic form factors. 
\end{abstract}

\maketitle

\section{Introduction}
\label{Sec:I}

The electromagnetic form factors (EMFFs) of baryons \cite{Bjorken:1966jh,Perdrisat:2006hj} play a crucial role in studying the internal structure of hadrons. The electric and magnetic formfactors of the baryons in the space-like region indicate the charge density and magnetic moment distributions, respectively, by Fourier transformation from the momentum space into the coordinate space.
In contrast, the physical meaning of the EMFFs in the timelike region is not so clear. There has been a long time effort trying to study the timelike EMFFs of the baryons, e.g., Refs.~\cite{Denig:2012by,Pacetti:2014jai} and references therein and thereafter.
Among them, Ref.~\cite{Yang:2022qoy}  proposed that the timelike EMFFs should reflect the distributions of the polarized charges and/or magnetic moment distributions, which is interesting but needs further study from both the theory and experimental sides. 
In the aspects of the experiment, the processes of electron-positron annihilation into baryon-anti-baryon pairs are rather powerful for studying the properties of the EMFFs \cite{Delcourt:1979ed,Antonelli:1998fv,Xia:2021agf,BaBar:2013ves,Achasov:2014ncd}.
Over the last decades, there have been many experimental measurements on nucleons as well as other baryons following this way. See, e.g., Refs.~\cite{Ablikim2018,Druzhinin:2019gpo,BESIII:2021rqk,BESIII:2021tbq,BESIII:2023ioy,BESIII:2023ynq}. These measurements strongly reduced the uncertainties compared with those of the old experiments.
Especially for almost all the data sets, if they give measurements around the baryon-anti-baryon thresholds, they find that there are strong enhancements very close to the thresholds.
This attracts enthusiastic attention from the theorists, such as works about $e^+e^-\to \bar{p}p$ \cite{Haidenbauer:2014kja,Bianconi:2015owa,Dai:2023vsw,Qian:2022whn,Rosini:2024lld}, $e^+e^-\to \bar{\Lambda}\Lambda$ \cite{Haidenbauer:2016won,Cao:2018kos,Xiao:2019qhl,Haidenbauer:2023zcu}, $e^+e^-\to \Sigma\bar{\Sigma}, \Xi\bar{\Xi}$  \cite{Haidenbauer:2020wyp,Yan:2023yff,Bai:2023dhc}, and $e^+e^-\to\Lambda_c^+\bar{\Lambda}_c^-$\cite{Chen:2023oqs, Dai:2017fwx}.

In the present analysis, we focus on the EMFFs of the $\Lambda_c$, which are somehow similar to that of the nucleons, with a $u/d$ valence quark replaced by the $c$ quark.
The cross-section of the reaction $e^+ e^-\to \Lambda_c^+\bar{\Lambda}_c^-$ \cite{Belle:2008xmh} was first measured by Belle collaboration in 2008, with large uncertainties.
Ten years later, the BESIII collaboration performed a measurement on the cross-section of this process \cite{Ablikim2018}. Although it constrains only four energy points, they are very close to the threshold (below 4.60~GeV), showing obvious enhancements around the threshold, too.
One of the most likely mechanisms leading to the enhancement is the final state interactions (FSI) \cite{Dai:2014zta,Dai:2016ytz,Yao:2020bxx},
where the baryon-anti-baryon rescatterings are taken into account.
The FSI can be implemented by a two-step procedure: The Lippmann-Schwinger (LS) equation is applied to describe the $\bar{B}B$  scattering amplitudes ~\cite{Kang:2013uia,Dai:2017ont}, where $B$ represents baryon; Then, one can construct the production amplitudes with $\bar{B}B$ final states based on the distorted wave Born appproximation (DWBA) method \cite{Haidenbauer:2014kja,Dai:2017fwx, Haidenbauer:2020wyp}.
This two-step procedure has proven to be a successful tool for predicting the amplitudes and describing the enhancement well in the energy region around the  $\bar{B}B$ thresholds.
Typical works dealing with the FSI of baryon-anti-baryon pairs in such a way can be found in the processes of $e^+ e^-\to \eta\Lambda\bar{\Lambda}$ \cite{Haidenbauer:2023zcu}, $e^+ e^-\to \Lambda\bar{\Lambda}, \Sigma\bar{\Sigma}, \Xi\bar{\Xi}$ \cite{Haidenbauer:2020wyp}, $J/\psi\to \bar{p}p\gamma, \eta'\pi\pi\gamma, 3(\pi^+\pi^-)\gamma, K_S^0K_S^0\eta\gamma$ \cite{Dai:2018tlc,Yang:2022kpm}.
Indeed, there are some other ways to deal with the FSI in a similar two-step procedure, where only the ways to get the kernel of the $\bar{B}B$ scattering are different and so on for constructing the production amplitudes. See, e.g., \cite{Milstein:2022tfx,Salnikov:2023ipo}.
Here, we will apply the two-step procedure to study the process of
$e^+e^-\to \Lambda_c^+\bar{\Lambda}_c^-$, taking into the FSI with the LS equation and DWBA method.
By fitting to the experimental data, the amplitudes are fixed.
Finally, one can discuss the enhancement around the $\Lambda_c^+\bar{\Lambda}_c^-$ threshold and predict the individual EMFFs of the $\Lambda_c$.

This paper is organized as follows:
In Sec.~\ref{Sec:II}, we introduce the calculation of the $ \Lambda_c^+ \bar{\Lambda}_c^-\to \Lambda_c^+ \bar{\Lambda}_c^-$ scattering amplitude that is solved by LS equation, and the production amplitude of the process of $e^+ e^-\to \Lambda_c^+ \bar{\Lambda}_c^-$ that is obtained by the DWBA method. In Sec.~\ref{Sec:III}, we fit our amplitudes to the experimental data sets and extract the individual EMFFs of the $\Lambda_c$. Both the modulus and the phases of the EMFFs are predicted, and their properties are discussed. Finally,  a brief summary is given in Sec.~\ref{Sec:IV}.

\section{Formalism}\label{Sec:II}
As discussed in the introduction, we apply the two-step procedure: first, the hadronic scattering amplitude is solved by the LS equation, then the production amplitude is obtained by the DWBA method.

\subsection{\texorpdfstring{$\Lambda_c^+\bar{\Lambda}_c^-$}{} scattering amplitude}
Following our previous work \cite{Dai:2017fwx},  the $\Lambda_c^+\bar{\Lambda}_c^-$ can be produced according to  $e^+e^-\to\gamma^*\to\Lambda_c^+\bar{\Lambda}_c^-\to\Lambda_c^+\bar{\Lambda}_c^-$, where one-photon transition dominates, and only the $^3S_1-^3D_1$ coupled partial waves of the $\Lambda_c^+\bar{\Lambda}_c^-$ system need to be considered.
These partial wave amplitudes can be solved by the LS equation,
\begin{eqnarray}\label{Eq1:LSE}
	T_{L'' L'}\!\!\!&&\!\!\!(p'',p';E)=V_{L''L'}(p'',p';E)\!+\!\sum_{L}\int_{0}^{\infty}\frac{dp\, p^2}{(2\pi)^3}\nonumber\\
	&&\!\!\!\!\!\!\!\!\times V_{L''L}(p'',p;E)\frac{1}{E\!-\!2E_p\!+\!i0^+}T_{LL'}(p,p';E)\,,
\end{eqnarray}
where $p'=|\vec{p}'|$ and $p''=|\vec{p}''|$ are the initial and final momenta in the center-of-mass frame (c.m.f.) of the $\Lambda_c^+\bar{\Lambda}_c^-$ system, respectively.  $E=\sqrt{s}$ is the energy in c.m.f., and energy of the intermediate state is $E_p=\sqrt{p^2+M^2_{\Lambda_c}}$, with $M_{\Lambda_c}$ the mass of $\Lambda_c$ baryon.
Since only the $^3S_1-^3D_1$ partial waves are involved, the orbital angular momenta should be $L,L',L''=0,2$.

The potentials for the $\Lambda_c^+\bar{\Lambda}_c^-$ system are constructed in the spirit of chiral effective field theory (ChEFT).
Here, we apply the $SU(3)$ ChEFT.
The relevant chiral effective Lagrangians are given as \cite{Yan:1992gz,Zou:2023},
\begin{eqnarray}\label{Eq:L}
	\mathcal{L}&= & \frac{1}{2}\langle\bar{B}_{\bar{3}}\left(i \fslash{D}-m_{\bar{3}}\right) B_{\bar{3}}\rangle+g_6\langle\bar{B}_{\bar{3}} u^\mu \gamma_5 \gamma_\mu B_{\bar{3}}\rangle \nonumber\\
	&+&\left[~g_2\langle\bar{B}_{\bar{3}} u^\mu \gamma_5 \gamma_\mu B_{6}\rangle+g_4\langle\bar{B}_{\bar{3}} u_\mu B^{*\mu}_{6}\rangle +h.c.~\right] ,
\end{eqnarray}
where $ B_{\bar{3}}$, $B_6$, and $B_6^{*\mu}$ are the antitriplet, sextets of spin-$1/2$ and of spin-$3/2$ fields of the charmed baryons, respectively. They are assembled in $3\times3$ anti-symmetric or symmetric matrices,
\begin{eqnarray}
	 B_{\bar{3}}&=&\left(\begin{array}{ccc}
				0          & \Lambda_c  & \Xi_c^{+} \\
				-\Lambda_c & 0          & \Xi_c^{0} \\
				-\Xi_c^{+} & -\Xi_c^{0} & 0
			\end{array}\right),\nonumber\\[3mm] 
	 B_6&=&\left(\begin{array}{ccc}
	\Sigma_c^{++} & \frac{1}{\sqrt{2}} \Sigma_c^+ & \frac{1}{\sqrt{2}} \Xi_c^{\prime+} \\
	\frac{1}{\sqrt{2}} \Sigma_c^+ & \Sigma_c^{0} & \frac{1}{\sqrt{2}} \Xi_c^{\prime0} \\
	\frac{1}{\sqrt{2}} \Xi_c^{\prime+} & \frac{1}{\sqrt{2}} \Xi_c^{\prime0} & \Omega_c
	\end{array}\right), \nonumber\\[3mm] 
	 B_6^*&=&\left(\begin{array}{ccc}
		\Sigma_c^{*++} & \frac{1}{\sqrt{2}} \Sigma_c^{*+} & \frac{1}{\sqrt{2}} \Xi_c^{\prime*+} \\
		\frac{1}{\sqrt{2}} \Sigma_c^{*+} & \Sigma_c^{*0} & \frac{1}{\sqrt{2}} \Xi_c^{\prime*0} \\
		\frac{1}{\sqrt{2}} \Xi_c^{\prime*+} & \frac{1}{\sqrt{2}} \Xi_c^{\prime*0} & \Omega_c^*
		\end{array}\right).
\end{eqnarray}
The covariant derivative is given as $D_\mu B=\partial_\mu B+\Gamma_\mu B+B \Gamma_\mu^T$,
where the chiral operators $\Gamma_\mu$ and $u_\mu$ are given as
\begin{eqnarray}
	\Gamma_\mu=\frac{1}{2}(u^\dagger\partial_\mu u+u\partial_\mu u^\dagger) \,, \nonumber\\
	u_\mu=i(u^\dagger\partial_\mu u-u\partial_\mu u^\dagger)\nonumber \,,
\end{eqnarray}
with $u^2(x)=U(x)$. $U(x)$ is given as 
\begin{eqnarray}
	U&=&\exp\left(\frac{\sqrt{2}i\Phi}{F_\pi}\right)\,,\,\,\,\,   \text{with}\,\,\nonumber\\
	\Phi&=&\left(\begin{array}{ccc}
		\frac{\pi^0}{\sqrt{2}}+\frac{\eta}{\sqrt{6}} & \pi^{+} & K^{+} \\
		\pi^{-} & -\frac{\pi^0}{\sqrt{2}}+\frac{\eta}{\sqrt{6}} & K^0 \\
		K^{-} & \bar{K}^0 & -\frac{2 \eta}{\sqrt{6}}
		\end{array}\right).
\end{eqnarray}
Here, one has $F_\pi=0.922$~GeV~\cite{ParticleDataGroup:2022pth}.

The isospin of $\Lambda_c$ is $I=0$. Thus, there is no contribution from one pion exchange, which is widely studied in the nucleon-anti-nucleon scatterings. For $\Lambda_c^+\bar{\Lambda}_c^-$ scattering, there is only one kind of diagram from the one pseudoscalar exchange (OPE), i.e., the $\eta$ exchange. 
At next-to-leading order (NLO), the contributions of two pseudoscalar exchanges (TPEs), except for the football diagrams, can be absorbed into the  contact term due to the large mass difference 
between $\Lambda_c$ and the flavor partners such as $\Sigma_c$ and $\Xi_c$, e.g., $\Delta M= M_{\Sigma_c}-M_{\Lambda_c}\simeq 167$~MeV and $\Delta M'= M_{\Xi_c}-M_{\Lambda_c}\simeq 181$~MeV.
Also, after expansion, one would find that there is only $\Lambda_c\bar{\Lambda}_c K\bar{K}$ vertex.
\begin{figure}[htp]
	\centering
	\includegraphics[width=0.99\linewidth]{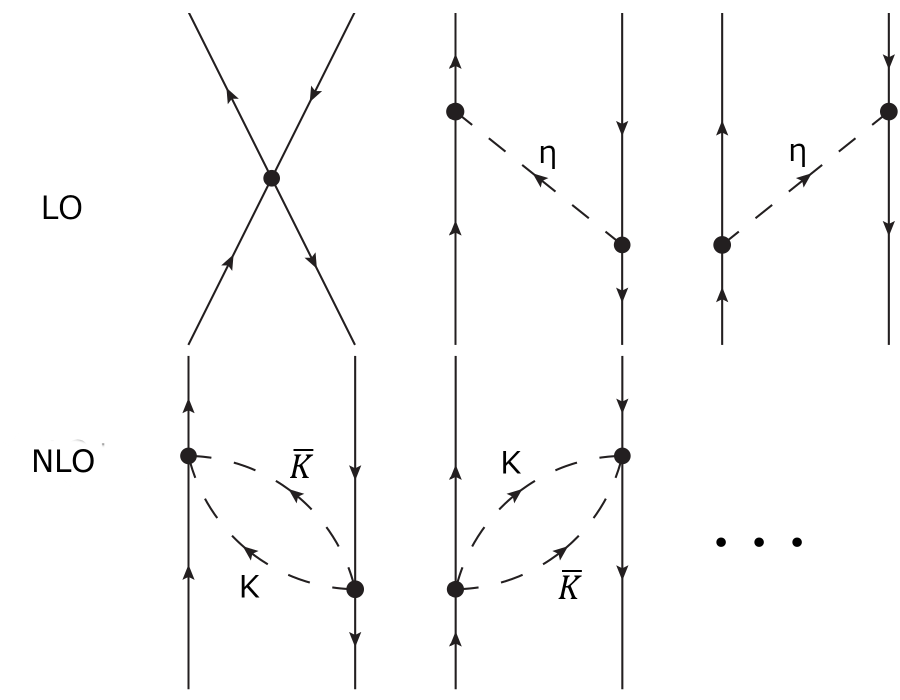}
	\caption{The Feynman diagram for $\Lambda_c \bar{\Lambda}_c$ scattering. The solid black line is $\Lambda_c$ and the dashed black line is $\eta$ or K.}
	\label{fig:Feynman}
\end{figure}
At the end of the day, one only needs to take into account the football diagrams of the $K\bar{K}$ exchanges, the contact and annihilation terms to get the potential up to NLO. The corresponding Feynman diagrams are shown in Fig.~\ref{fig:Feynman}.
After the expansion, the detailed interaction Lagrangians of the charmed baryons coupling with the pseudoscalar(s) that are necessary to calculate these diagrams are,
\begin{eqnarray}
\mathcal{L}^{\rm OPE}&=&-\frac{2 g_6 }{\sqrt{3}F_\pi}\bar{\Lambda }_c^-\gamma_5\gamma_\mu\Lambda _c^+\partial ^{\mu} \eta \nonumber\\
\mathcal{L}^{\rm TPE}&=&\frac{i}{4 F_\pi^2}\bar{\Lambda }_c^-\gamma_\mu\Lambda _c^+\big[K^0\partial^{\mu }\bar{K}^0-\bar{K}^0\partial^{\mu} K^0\nonumber\\
&&+K^+\partial^{\mu }K^--K^-\partial^{\mu} K^+\big]\,. 
\end{eqnarray}
It is worth pointing out that there are no unknown couplings for the football vertex, which can be used as the calibration to compare the strength of other diagrams.
The potential with OPE and TPE up to NLO is given as
\begin{eqnarray}\label{vobe}
	V_{\Lambda_c \bar{\Lambda}_c \rightarrow \Lambda_c \bar{\Lambda}_c}^{\rm OPE+TPE}&=&V_{\sigma q}\left(\vec{\sigma}_1 \cdot \vec{q}\right)\left(\vec{\sigma}_2 \cdot \vec{q}\right)+V_C \,,
\end{eqnarray}
where $\vec{q}$ is the transfer momentum, $\vec{q}=\vec{p}'-\vec{p}$. $V_{\sigma q}$ is from OPE, 
\begin{equation}
	V_{\sigma q}=-\frac{4g_6^2}{3F_\pi^2}\frac{1}{\vec{q}^2+m_\eta^2} \,,
\end{equation}
and $V_C$ is from TPE, i.e., the football diagrams,   
\begin{eqnarray}
	V_{C}&=&\frac{-w^2(q)L(q)}{192\pi^2F_\pi^4} \,,
\end{eqnarray}
where one has 
\begin{eqnarray}
w(q)&=&\sqrt{(q^2+4m_K^2)}, \nonumber\\
L(q)&=&\frac{w(q)}{q} \ln \frac{w(q)+q}{2  m_K }\,.
\end{eqnarray}
Transforming the potentials into the $LSJ$ representation, one has
\begin{eqnarray}\label{Eq:3S1-3D1}
	V^{\rm OPE+TPE}_{^3S_1}(p',p) \!&=&\!\langle 0,11|V_{\Lambda_c \bar{\Lambda}_c}|0,11\rangle  \nonumber\\
	\!&=&\!2\pi\!\int^1_{-1}\!dz\bigg[
	\frac{(p'^2\!+\!p^2)V_{\sigma q}+3V_C}{3}P_{0}(z)  \nonumber\\
	&&\;\;\;-\frac{2p'pV_{\sigma q}}{3}P_{1}(z)\bigg],\nonumber\\
	V^{\rm OPE+TPE}_{^3D_1}(p',p) \!&=&\!
	\langle 2,11|V_{\Lambda_c \bar{\Lambda}_c}|2,11\rangle\nonumber\\
	\!&=&\!2\pi\int^1_{-1}dz\bigg[-\frac{2p'pV_{\sigma q}}{3}P_{1}(z)  \nonumber\\
	&&\;\;\;+\frac{3V_C-(p'^2+p^2)V_{\sigma q}}{3}P_{2}(z)\bigg]\,,\nonumber\\
	V^{\rm OPE+TPE}_{^3D_1-^3S_1}(p',p)\!&=&\!\langle 2,11|V_{\Lambda_c \bar{\Lambda}_c}|0,11\rangle\nonumber\\
	\!&=&\!2\pi\sqrt{\frac{2}{3}}\int^1_{-1}dz\bigg[-4p'p V_{\sigma q}P_{1}(z)
	\nonumber\\
	&+&\!2p'^2V_{\sigma q}P_{0}(z)+2p^2V_{\sigma q}P_{2}(z)\bigg]\,,\nonumber\\
	V^{\rm OPE+TPE}_{^3S_1-^3D_1}(p',p)\!&=&\!\langle 0,11|V_{\Lambda_c \bar{\Lambda}_c}|2,11\rangle\nonumber\\
	\!&=&\!2\pi\sqrt{\frac{2}{3}}\int^1_{-1}dz\bigg[-4p'p V_{\sigma q}P_{1}(z) \nonumber\\
	 &+&\!2p'^2V_{\sigma q}P_{2}(z)
	+2p^2V_{\sigma q}P_{0}(z)\bigg].
\end{eqnarray}
In Ref.\cite{Yan:1992gz}, they obtained $g_6=0$ by comparing with the one pion emission matrix element. Here, we will follow this constraint. Indeed, as we have been tested, the $g_6$ will be very small in the fit, and the results will not be changed much by adding/removing the $\eta$ exchange contribution. Hence, we can safely ignore the contribution of OPE in the present analysis.

The contact and annihilation potentials for $^3S_1-^3D_1$ coupled partial waves can be written explicitly as \cite{Dai:2017ont}
\begin{eqnarray}
	V_{^3S_1}(p',p)\!&=&\!\tilde{C}_{^3S_1}\!+\!C_{^3S_1}(p'^2\!+\!p^2)\nonumber\\
	\!&-&\!i(\tilde{C}^a_{^3S_1}\!+\!C^a_{^3S_1}p'^2)(\tilde{C}^a_{^3S_1}\!+\!C^a_{^3S_1}p^2)\,,\nonumber\\
	V_{^3D_1-^3S_1}(p',p)\!&=&\!C_{\epsilon_1}p'^2\!-\!iC^a_{\epsilon_1}p'^2(\tilde{C}^a_{^3S_1}\!+\!C^a_{^3S_1}p^2)\,,\nonumber\\
	V_{^3S_1-^3D_1}(p',p)\!&=&\!C_{\epsilon_1}p^2\!-\!iC^a_{\epsilon_1}p^2(\tilde{C}^a_{^3S_1}\!+\!C^a_{^3S_1}p'^2)\,,\nonumber\\
	V_{^3D_1}(p',p)\!&=&0\!\,,
\end{eqnarray}
where $\tilde{C}_i$, $C_i$ and $\tilde{C}^a_i$, $C^a_i$ are low-energy constants (LECs) of the contact and annihilation potentials, respectively. They are all real numbers. Notice that at leading order (LO), only $\tilde{C}_{^3S_1}$ and $\tilde{C}^a_{^3S_1}$ need to be included. 
A regulator function is necessary to be multiplied on the potentials to suppress the ultraviolet divergence in the high-energy region caused by the momenta. Following the previous work \cite{Dai:2017fwx}, an explicit form for the regulator is given as follows
\begin{eqnarray}\label{Eq3:regulator}
	f(p',p)=\exp\left(-\frac{p'^2+p^2}{\Lambda^2}\right)\,,
\end{eqnarray}
where the cutoff $\Lambda$ is chosen as 450-650~MeV, with the interval of 50~MeV.

\subsection{The production amplitude and observables}
With the hadronic scattering amplitudes calculated in the previous section, the $\gamma \Lambda_c^+\bar{\Lambda}_c^-$ form factor can be obtained by the DWBA method,
\begin{eqnarray}\label{Eq4:DWBA}
	f_L^{\Lambda_c^+\bar{\Lambda}_c^-}(p)\!&=&\!f_L^{\Lambda_c^+\bar{\Lambda}_c^-,0}(p)\!+\!\sum_{L'}\int_0^{\infty}\frac{dp'\,p'^2}{(2\pi)^3}f_{L'}^{\Lambda_c^+\bar{\Lambda}_c^-,0}(p')\nonumber\\
	&&\!\!\!\times\frac{1}{E-2E_{p'}+i0^+}T_{L'L}(p,p';E)\,,
\end{eqnarray}
where $E=\sqrt{s}=2\sqrt{M_{\Lambda_c}^2+p^2}$ is the c.m.f. energy.
The relation between $f_L^{\Lambda_c^+\bar{\Lambda}_c^-}$ and the EMFFs, $G_{M}^{\Lambda_c}$ and $G_E^{\Lambda_c}$, are given as follows
\begin{eqnarray}\label{Eq5:gammaLcLcbar}
	f_0^{\Lambda_c^+\bar{\Lambda}_c^-}(p)\!&=&\!G_M^{\Lambda_c}(p)+\frac{M_{\Lambda_c}}{\sqrt{s}}G_E^{\Lambda_c}(p)\,,\nonumber\\
	f_2^{\Lambda_c^+\bar{\Lambda}_c^-}(p)\!&=&\!\frac{1}{\sqrt{2}}\left(G_M^{\Lambda_c}(p)-\frac{2M_{\Lambda_c}}{\sqrt{s}}G_E^{\Lambda_c}(p)\right)\,.
\end{eqnarray}
Once the $\gamma \Lambda_c^+\bar{\Lambda}_c^-$ form factors are fixed, the $G_{M}^{\Lambda_c}$ and $G_E^{\Lambda_c}$ can be extracted from Eq.\eqref{Eq5:gammaLcLcbar}.
The Born term, i.e., the bare $\gamma \Lambda_c^+\bar{\Lambda}_c^-$ vertex without $\Lambda_c^+\bar{\Lambda}_c^-$ rescattering,  $f_L^{\Lambda_c^+\bar{\Lambda}_c^-,0}$, can be parameterized as
\begin{eqnarray}\label{Eq5:gammaLcLcbar;Born}
	f_0^{\Lambda_c^+\bar{\Lambda}_c^-,0}(p)\!&=&\!G_M^{\Lambda_c,0}+\frac{M_{\Lambda_c}}{\sqrt{s}}G_E^{\Lambda_c,0}\,,\nonumber\\
	f_2^{\Lambda_c^+\bar{\Lambda}_c^-,0}(p)\!&=&\!\frac{1}{\sqrt{2}}\left(G_M^{\Lambda_c,0}-\frac{2M_{\Lambda_c}}{\sqrt{s}}G_E^{\Lambda_c,0}\right)\,.
\end{eqnarray}
In general, the bare EMFFs, $G_{E,M}^{\Lambda_c,0}$, should have momentum dependence and can be complex. However, in practice, real constants are good enough to fit the experimental data. See discussions in the next section.
Also, people must obey the condition, $G_M^{\Lambda_c,0}=G_E^{\Lambda_c,0}$. It is because the electric and magnetic form factors should be the same at the threshold, as required by the definition of the Sachs form factors \cite{Sachs:1962zzc}.
With the form factor of the $\gamma \Lambda_c^+\bar{\Lambda}_c^-$ vertex, one can build the partial wave amplitude of the reaction $e^+ e^-\to \Lambda_c^+\bar{\Lambda}_c^-$. It is
\begin{eqnarray}
	F_{LL'}^{\Lambda_c^+\bar{\Lambda}_c^-,e^+e^-}=-\frac{4\alpha}{9}	f_L^{\Lambda_c^+\bar{\Lambda}_c^-}f_{L'}^{e^+e^-}\,.
\end{eqnarray}
where $f_{L'}^{e^+e^-}$ is the $\gamma e^+ e^-$ vertex,
\begin{eqnarray}
	f_0^{e^+e^-}&=&1+\frac{m_e}{\sqrt{s}}\,,\nonumber\\
	f_2^{e^+e^-}&=&\frac{1}{\sqrt{2}}\left(1-\frac{2m_e}{\sqrt{s}}\right)\,.
\end{eqnarray}
Finally, one can get the cross-section of $e^+ e^-\to \Lambda_c^+\bar{\Lambda}_c^-$ based on the partial wave amplitudes obtained above,
\begin{eqnarray}
	\sigma(s)
	&=&\frac{3\pi \beta}{s}C(s)\left(|F_{00}^{\Lambda_c^+\bar{\Lambda}_c^-,e^+e^-}|^2+|F_{02}^{\Lambda_c^+\bar{\Lambda}_c^-,e^+e^-}|^2\right.\nonumber\\
	&&\;\;+\left.|F_{20}^{\Lambda_c^+\bar{\Lambda}_c^-,e^+e^-}|^2+|F_{22}^{\Lambda_c^+\bar{\Lambda}_c^-,e^+e^-}|^2\right)\nonumber\\
	&=&\frac{4\pi \alpha^2\beta C(s)}{3s}\left[|G_M^{\Lambda_c}|^2+\frac{2 M_{\Lambda_c}^2}{s}|G_E^{\Lambda_c}|^2\right]\,, \label{Eq:cs}
\end{eqnarray}
where one has $\beta=k_{\Lambda_c}/k_e$, with $k_{\Lambda_c}(k_e)$ the momenta in c.m.f. of $\Lambda_c^+\bar{\Lambda}_c^-$ ($e^+ e^-$) system. The S-wave Sommerfeld-Gamow factor is $C(y)=y/(1-e^{-y})$ with $y=\pi \alpha M_{\Lambda_c}/k_{\Lambda_c}$. The fine-structure constant is  $\alpha=1/137.036$.
According to Eq.~(\ref{Eq:cs}), the effective form factor can also be extracted, which is defined as 
\begin{eqnarray}\label{Eq:effect;EMFF}
|G_{eff}|=\sqrt{\frac{\sigma_{e^+ e^-\to \Lambda_c^+\bar{\Lambda}_c^-}}{\frac{4\pi \alpha^2\beta}{3s}C(s)\left(1+\frac{2M_{\Lambda_c}}{s}\right)}}\,.
\end{eqnarray}
As will be discussed in the next section, the effective EMFFs help to study the contributions from different kinds of potentials.

\section{Fit Results and Discussion}
\label{Sec:III}
\subsection{Fit results}
With amplitudes discussed above, one can perform LO and NLO fits on data from BESIII collaboration, including the cross-section of $e^+e^-\to\Lambda_c^+\bar{\Lambda}_c^-$ and the modulus of the individual EMFFs as well as their ratios.
For the LO analysis, only three parameters need to be fixed, that is, $\tilde{C}_{^3S_1}$, $\tilde{C}^a_{^3S_1}$, and the bare EMFFs $G_{M}^{\Lambda_c,0}(=G_{E}^{\Lambda_c,0})$. 
In principle, the bare EMFFs can be complex numbers that are free to be fixed by the fit. Also, they can have momentum dependence. However, in practice, one can set them as a real constant, with little loss of the quality of the solution. 
Moreover, it is found that the bare EMFFs will be very close to one by fitting the data. Hence, we fixed them as $G_{M}^{\Lambda_c,0}=G_{E}^{\Lambda_c,0}=1$. 
For the NLO fit, there are six parameters that need to be determined. Other than the two parameters used in the LO fit, one should also include $C_{^3S_1}$, $C^a_{^3S_1}$, $C_{\epsilon_1}$ and $C^a_{\epsilon_1}$ for the contact and annihilation potentials.  
Besides, we consider a series of cutoffs, $\Lambda=450,\,500,\,550,\,600,\, 650$ MeV in the 
regulator to test the dependence on cutoff. See Eq.\eqref{Eq3:regulator}. 
High-quality solutions are obtained, and the fit values of the parameters are listed in Table \ref{Table1}. 
\begin{table}[htp]
\centering
\begin{tabular}{ccccccc}
	\hline
	                                  & LO     & \multicolumn{5}{c}{NLO}                                     \\ \hline
	$\Lambda$                         & 500    & 450                     & 500    & 550    & 600    & 650    \\
	$\tilde{C}_{^3S_1}$(GeV$^{-2}$)   & 0.006  & 0.017                   & -0.013 & -0.031 & -0.038 & -0.027 \\
	$C_{^3S_1}$(GeV$^{-4}$)           & --     & 0.215                   & 0.177  & 0.1615 & 0.143  & 0.137  \\
	$C_{\epsilon_1}$(GeV$^{-4}$)      & --     & 0.080                   & -0.087 & 0.008  & 0.045  & 0.064  \\
	$\tilde{C}^a_{^3S_1}$(GeV$^{-1}$) & -0.140 & -0.012                  & -0.100 & -0.180 & -0.200 & -0.178 \\
	$C^a_{^3S_1}$(GeV$^{-3}$)         & --     & 0.030                   & 0.651  & 0.879  & 0.876  & 0.856  \\
	$C^a_{\epsilon_1}$(GeV$^{-3}$)    & --     & -9.999                  & -1.601 & -0.458 & -0.170 & -0.019 \\
	$\chi^2/{\rm d.o.f.}$             & 2.072  & 1.108                   & 0.364  & 0.428  & 0.493  & 0.538  \\ \hline
\end{tabular}
\caption{The LECs and bare EMFF of our solutions. Notice that we fix some of the parameters as $G_{M,E}^{\Lambda_c,0}=1$ and $g_6=0$.}
\label{Table1}
\end{table}
The fit is performed within MINUIT \cite{James:1975dr}. 
As can be found, the $\chi^2/{\rm d.o.f.}$ of NLO, except for that with $\Lambda=450$~MeV, is even smaller than 0.6,  indicating how well the solutions fit the data! Among them, the one with cutoff $\Lambda=500$~MeV has the smallest $\chi^2/{\rm d.o.f.}$. Also, as will be discussed in the next paragraphs, it describes the data best, so we choose it as the optimal solution. 

\begin{figure}[htp]
	\centering
	\includegraphics[width=0.99\linewidth,height=0.5\textheight]{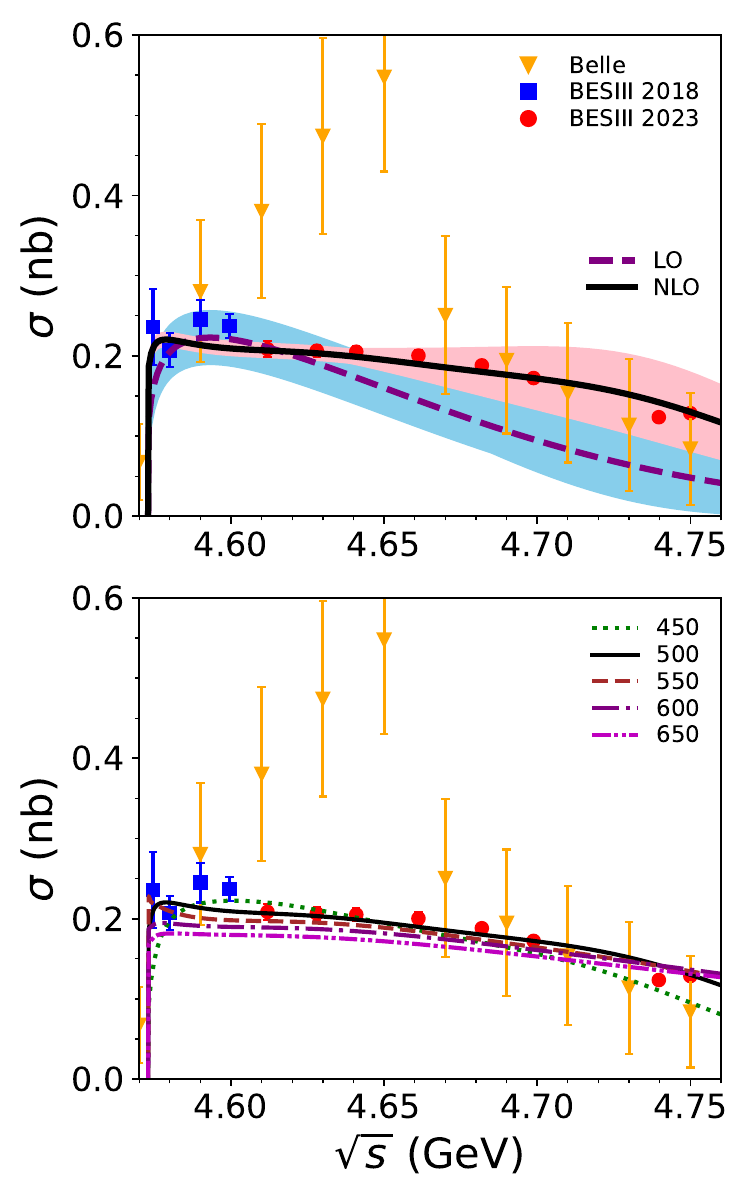}
	\caption{Comparison between our results and the cross-sections from the experiment. The data sets are taken from Belle \cite{Belle:2008xmh} and BESIII \cite{Ablikim2018,Ablikim2023} collaborations. The LO and NLO results with cutoff $\Lambda$=500 MeV are shown at the top, and that of NLO results with cutoffs $\Lambda$=450, 500, 550, 600, 650 MeV are shown at the bottom.}
	\label{Fig:Sigma}
\end{figure}

The fit results to the cross-sections are shown in Fig.~\ref{Fig:Sigma}.
In the fitting, only the experimental data sets from BESIII collaboration \cite{Ablikim2018,Ablikim2023} are taken into account, while that of the Belle collaboration \cite{Belle:2008xmh} is only plotted for the reader's convenience due to their large uncertainties. 
The top graph gives results at LO (purple dashed line) and NLO (black solid line) with cutoff $\Lambda$=500 MeV. 
Their uncertainties are shown as sky blue and pink bands for LO and NLO, respectively. They are estimated from a Bayesian method following Refs.~\cite{Epelbaum:2014efa,Dai:2017ont}.   
For example, 
the uncertainty $\Delta X^{\mathrm{NLO}}(p)$ of the NLO observable  $X^{\mathrm{NLO}}(p)$ is estimated by
\begin{equation} \label{eq:error}
    \Delta X^{\mathrm{NLO}}(p)\!=\!\max\left(Q^3|X^{\mathrm{LO}}(p)|, Q|X^{\mathrm{LO}}(p)\!-\!X^{\mathrm{NLO}}(p)|\right),
\end{equation}
with the non-dimensional parameter $Q$ defined as
\begin{equation}
    Q=\max\left(\frac{p}{\Lambda_b},\frac{M_\pi}{\Lambda_b}\right), \nonumber
\end{equation}
where $\Lambda_b$ is the breakdown scale, and we can set $\Lambda_b=900$ MeV in the present analysis. The observable $X(p)$ can be either the cross-section or the individual EMFFs. 
Though this uncertainty estimation does not provide a statistical interpretation, it gives a more reliable estimation than relying on cutoff variations and is successful in the phenomenology study within ChEFT \cite{Dai:2017ont,Yang:2022kpm}.
As can be found, both solutions describe the data well below 4.63~GeV ($p=362$~MeV). Nevertheless, the NLO one is still consistent with the data in the higher energy region, but the LO one is not. This is compatible with the characteristic of the ChEFT, where the NLO amplitudes work better than the LO ones and can extend the working energy region.    

The bottom graph gives results at NLO with different cutoffs, $\Lambda$=450, 500, 550, 600, 650 MeV, shown as green dotted, black solid, brown dashed, purple dash-dotted, and magenta dash-dot-dotted lines. As can be found, the cross-section has a strong enhancement that is very close to the threshold and then keeps flat up to 4.65 GeV. Then, it decreases gradually in the energy region from 4.65-4.75 GeV. 
This confirms the threshold enhancement discovered in the cross sections of many processes of electron-positron annihilation into baryon-anti-baryon pairs, e.g., $e^+e^-\to\bar{N}N$ \cite{Haidenbauer:2014kja,Yang:2022qoy}, $e^+e^-\to\bar{\Lambda}\Lambda$ \cite{Haidenbauer:2016won,Cao:2018kos}. 
Among the fits with different cutoffs, the solution with  $\Lambda=500$~MeV describes the data best from an overall point of view. The one with $\Lambda=450$~MeV is not so consistent with the first data point, while the other ones with cutoffs $\Lambda=550, 600,650$~MeV are worse in the high energy data points shown as the red circles. Hence, we take $\Lambda=500$~MeV as the optimal solution. 
Nevertheless, one can still find that all curves with different cutoffs almost overlap with each other, with only slight differences, implying the reliability of our model.  

\begin{figure}[htp]
	\centering
	\includegraphics[width=0.99\linewidth, height=0.3\textheight]{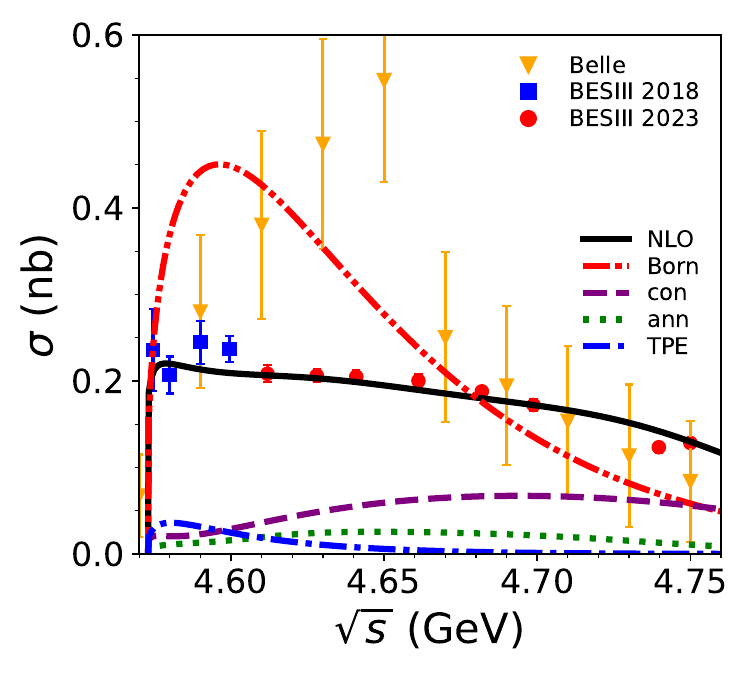}
	\caption{Separated contributions to the cross-section. The black solid, red dash-dot-dotted, purple dashed, blue dash-dotted, and green dotted lines are for the total, Born term, contact, annihilation, and TPE contributions, respectively. The data is from Belle \cite{Belle:2008xmh} and BESIII \cite{Ablikim2018,Ablikim2023} collaborations. The cutoff is chosen as $\Lambda$=500~MeV.}
	\label{Fig:component}
\end{figure}
To study the contributions to the cross sections from each kind of potential, we list the separated contributions from the Born term, contact, annihilation, and TPE potentials. See Fig.~\ref{Fig:component}. 
As can be seen, the Born term, as shown by the red dash-dot-dotted line, dominates in the whole energy region. However, the Bron term is far away from the total contribution. Thus, 
complicated interferences between it and all other components, the contact, annihilation, and TPE potentials are necessary. 
Besides, the TPE contribution is larger than the contact and annihilation ones in the low-energy region, and the contact terms contribute more in the high-energy region. This is compatible with the property of ChEFT, where the contact term contributes significantly in the short-distance region, and the pseudoscalar exchanges contribute much in the long-range region. 
Also, as will be discussed in the following subsection, the contact potentials will make significant contributions in the energy region that is very close to the threshold.

\begin{figure}[htp]
	\centering
	\includegraphics[width=0.99\linewidth,height=0.16\textheight]{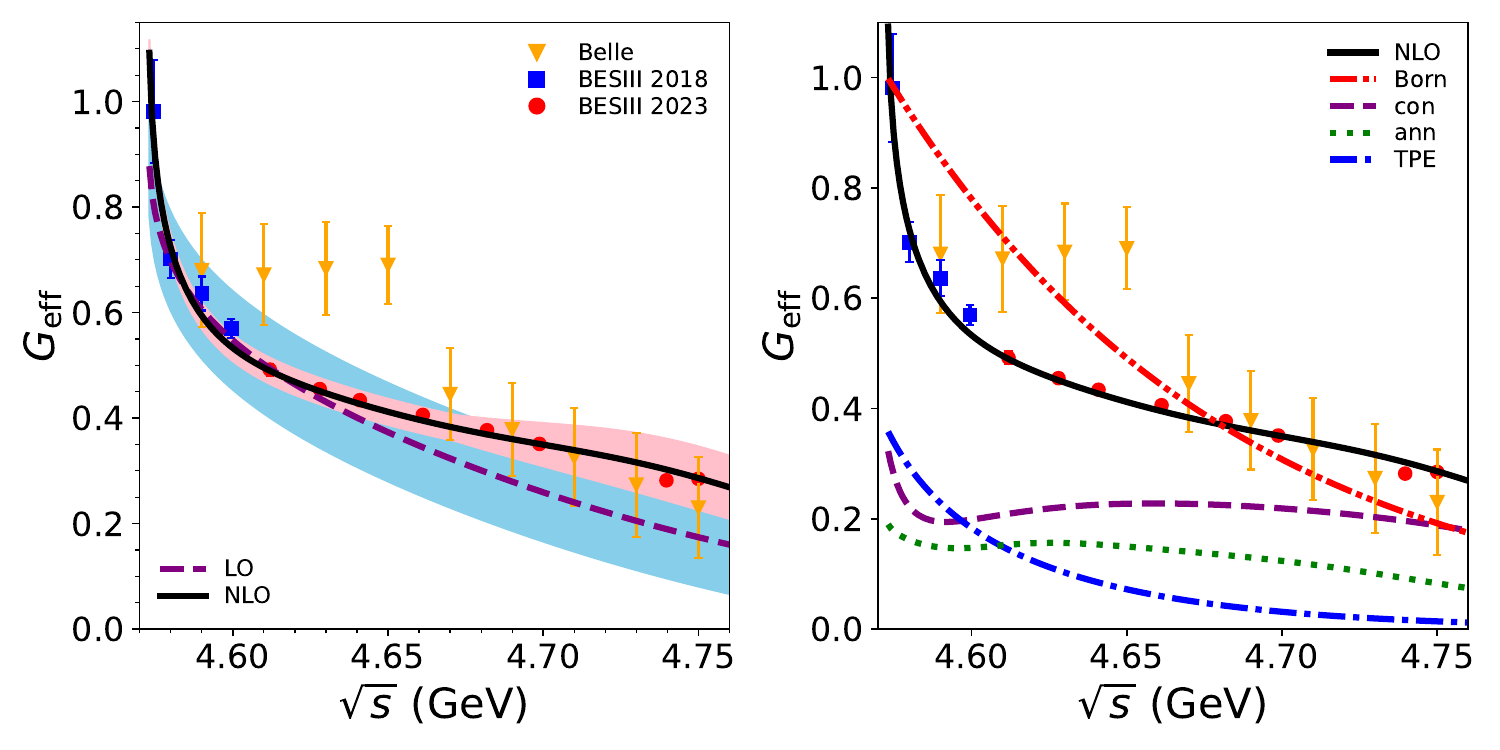}
	\caption{Comparison of the effective EMFFs between ours and the experimental data. 
The data sets are taken from Belle \cite{Belle:2008xmh} and BESIII \cite{Ablikim2018,Ablikim2023} collaborations. 
The LO and NLO results are shown in the left column, and the NLO separated contributions from the total, Born term, contact, annihilation, and TPE potentials are shown in the right column. The cutoff is fixed as  $\Lambda$=500~MeV.}
	\label{Fig:Geff}
\end{figure}
According to Eq.\eqref{Eq:effect;EMFF}, one can calculate the effective EMFFs of the $\Lambda_c$ from the cross-section of the reaction $e^+ e^-\to \Lambda_c^+\bar{\Lambda}_c^-$. 
The Comparison between our results and the experimental data is shown in Fig.\ref{Fig:Geff}. 
As can be seen in the left graph, our results fit the data well. This is not a surprise, as the effective EMFFs are related to the cross-section. See Eq.~(\ref{Eq:effect;EMFF}). 
It is found that the effective EMFF decreases rapidly with an increase in energy, and then it decreases slowly. There are even some tiny fluctuations in the high-energy region that are suppressed by momentum. 
This kind of behavior is very similar to that of the nucleon \cite{Yang:2022qoy}.
Moreover, now the contribution of the contact term is more apparent due to the fact that the effective EMFFs are obtained by dividing out the momentum factor $k_{\Lambda_c}$ from the cross-section, as shown in Eq.~(\ref{Eq:effect;EMFF}). 
Indeed, the contact potential leads to an enhancement near the threshold. See the purple dashed line in Fig.~\ref{Fig:Geff}. 
As has been tested, if we remove the contact term, the cross-section can not be fitted well, especially in the low-energy region.

\subsection{Individual EMFFs}
\begin{figure}[htp]
	\centering
	\includegraphics[width=0.99\linewidth,height=0.48\textheight]{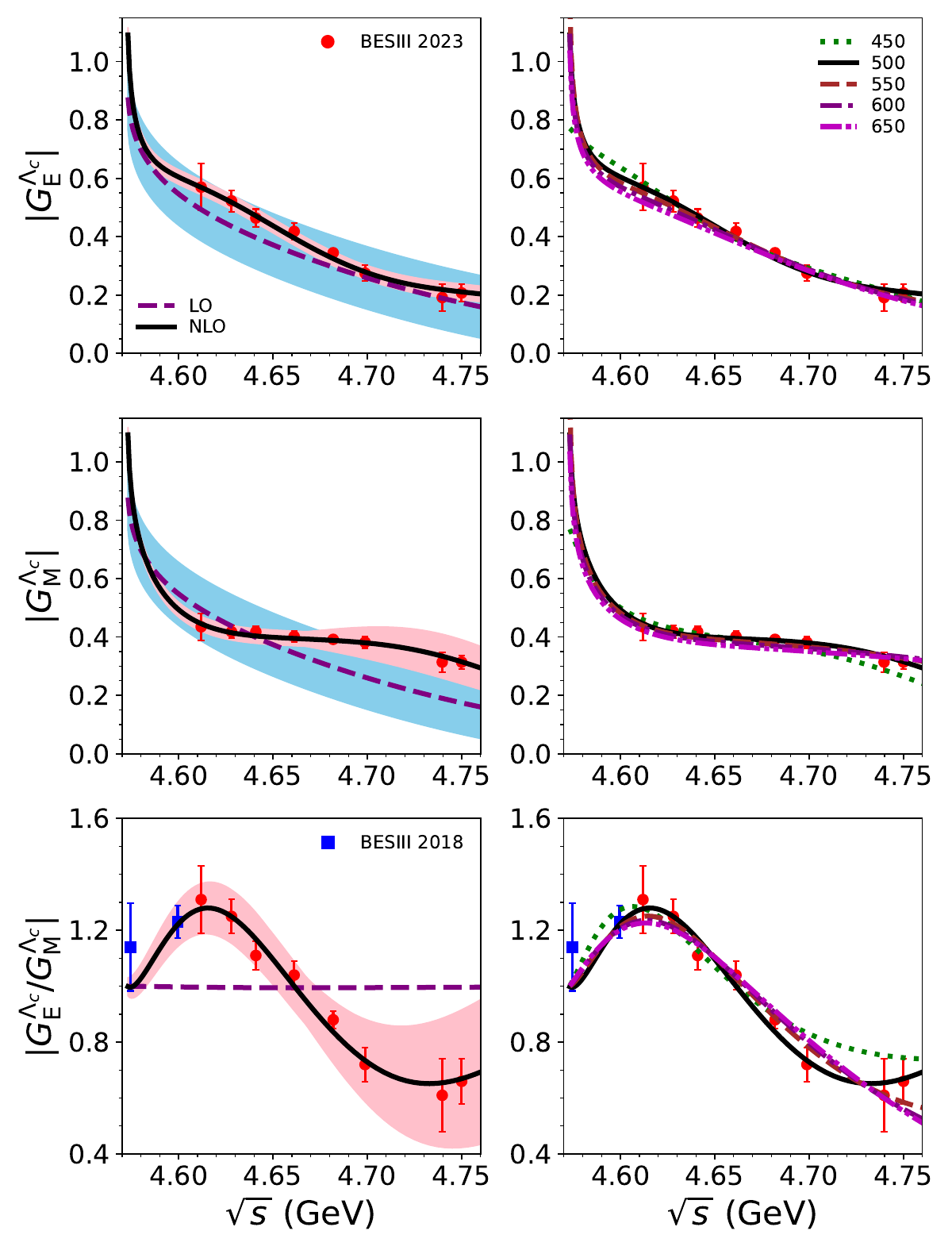}
	\caption{Our results of the modulus of the individual EMFFs, $|G_E^{\Lambda_c}|$, $|G_M^{\Lambda_c}|$ and their ratio,  $|G_E^{\Lambda_c}/G_M^{\Lambda_c}|$. The data are from BESIII collaboration \cite{Ablikim2018,Ablikim2023}. The LO and NLO results are shown in the left column, with cutoff $\Lambda=500$~MeV. The NLO results with different cutoffs are shown in the right column.  }
	\label{Fig:gme}
\end{figure}

With the amplitudes obtained in the last section, we can extract the individual EMFFs, $G_M^{\Lambda_c}$ and $G_E^{\Lambda_c}$. See Eq.\eqref{Eq5:gammaLcLcbar}. 
The results are shown in Fig. \ref{Fig:gme}. 
In the left column, the graphs are for results at LO and NLO, with cutoff $\Lambda$=500 MeV. In the right column, they are results at NLO with different cutoffs, $\Lambda$=450, 500, 550, 600, 650 MeV.  
The modulus of the electric and magnetic form factors, as well as 
their ratio ($G_E^{\Lambda_c}$, $G_M^{\Lambda_c}$, $G_E^{\Lambda_c}/G_M^{\Lambda_c}$) are shown in the top, middle, and bottom rows of Fig.~\ref{Fig:gme} in order. 
The $|G_E^{\Lambda_c}|$ and $|G_M^{\Lambda_c}|$ at LO are almost the same. See discussions on the ratio ($|G_E^{\Lambda_c}/G_M^{\Lambda_c}|$) below. 
The NLO results are pretty consistent with the BESIII data, while the LO results are much worse. This is compatible with the prediction ability of the ChEFT at different orders.   
For the LO results, the fit of electric and magnetic form factors is better than that of the ratio ($|G_E^{\Lambda_c}/G_M^{\Lambda_c}|$), where the solution is not far away from the first four data points of the electric and magnetic form factors, but only close to one data point of the ratio.  
This indicates that the experimental measurement of the ratio is quite crucial for refining theoretical analysis.

The individual EMFFs decrease rapidly in the increase of energy in the region close to the threshold, and then they decrease much slower in the high-energy region, where even some fluctuations can be found. 
Specifically, there is a difference between the variation of $|G_E^{\Lambda_c}|$ and $|G_M^{\Lambda_c}|$ at NLO. Compared with $|G_E^{\Lambda_c}|$, $|G_M^{\Lambda_c}|$ decreases more rapidly not far away from the threshold. This is reflected by the fact that $|G_E^{\Lambda_c}/G_M^{\Lambda_c}|$ increases from 1 to 1.3 in the energy region from threshold to 4.61 GeV. 
Then both of them decrease slower in the high energy region, but $|G_E^{\Lambda_c}|$ has more fluctuations than $|G_M^{\Lambda_c}|$, leading to the fluctuations of  $|G_E^{\Lambda_c}/G_M^{\Lambda_c}|$, too.

For the results of the ratio $|G_E^{\Lambda_c}/G_M^{\Lambda_c}|$, as can be seen in the bottom two graphs, the NLO result is consistent with the BESIII data and has oscillation behavior. 
The LO result is almost flat. This is because only the $^3S_1$ wave of the contact and annihilation potentials is left at LO. As a result, the Born term dominates in the D-wave production amplitude, $f_2^{\Lambda_c^+\bar{\Lambda}_c^-}$. See Eq.~(\ref{Eq4:DWBA}). Correspondingly, $f_2^{\Lambda_c^+\bar{\Lambda}_c^-}$ would be small as $G_{E}^{\Lambda_c, 0}=G_{M}^{\Lambda_c, 0}$. Finally, one roughly has $G_E^{\Lambda_c}/G_M^{\Lambda_c}\simeq \sqrt{s}/2M_{\Lambda_c}\simeq 1$ in the assumption of small momentum.    

\begin{figure}[htp]
	\centering
	\includegraphics[width=0.99\linewidth,height=0.48\textheight]{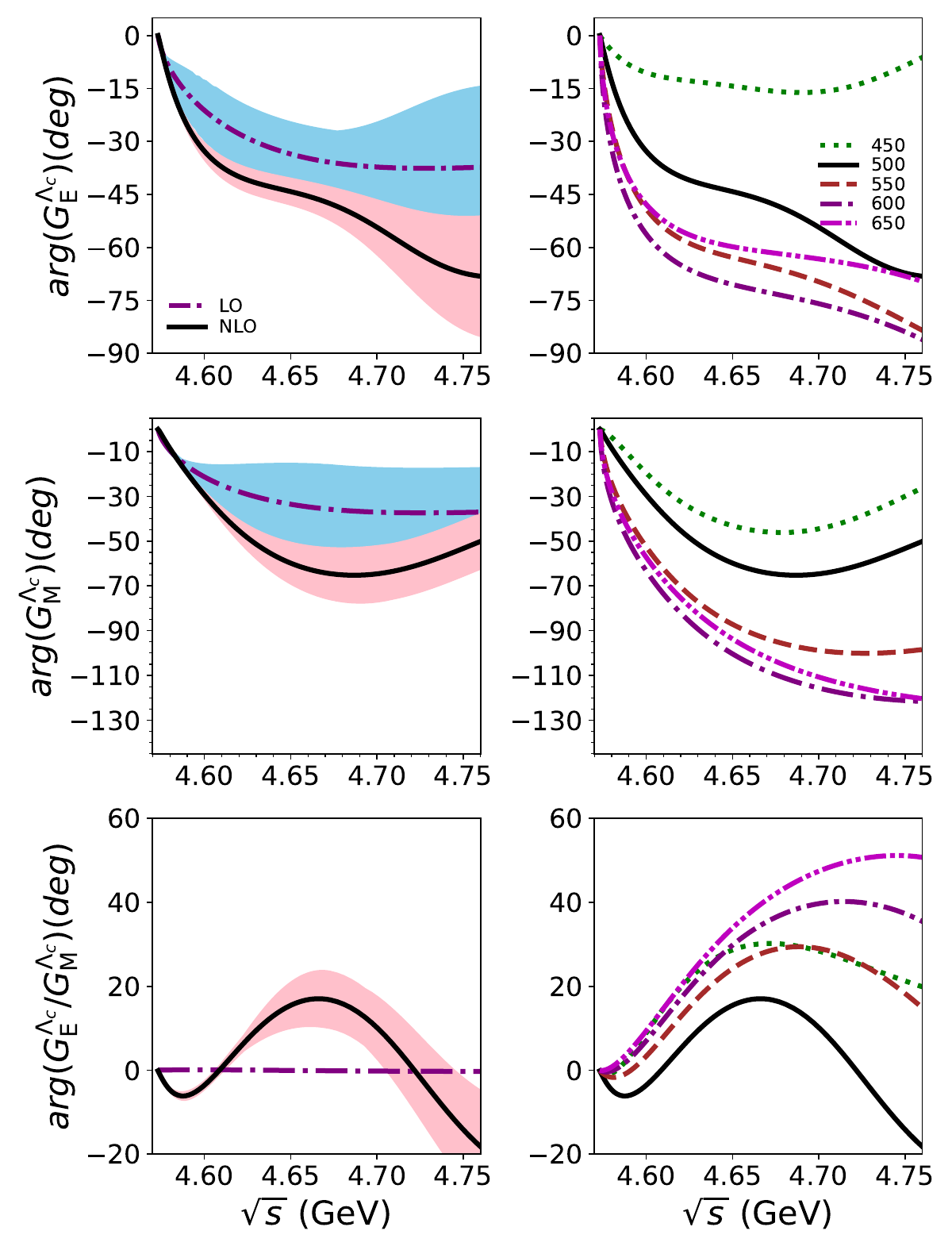}
	\caption{Our predictions of the phases of individual EMFFs, $arg(G_E^{\Lambda_c})$, $arg(G_M^{\Lambda_c})$, and $arg(G_E^{\Lambda_c}/G_M^{\Lambda_c})$. The LO and NLO results are shown in the left column, with cutoff $\Lambda=500$~MeV. The NLO results with different cutoffs are shown in the right column. }
	\label{Fig:gmean}
\end{figure}

The NLO results with cutoffs $\Lambda$=450, 500, 550, 600, 650 MeV are shown in the graphs in the right column of Fig.~\ref{Fig:gme}. The curves almost overlap with each other except for those of the tails in the high-energy region. This is not strange, as the tails are close to the edge of the working energy region of our model. From an overall point of view, our solutions fit the data rather well, indicating again that our results are not sensitive to the effect of cutoffs, confirming the stability of our model.

Besides, we also predict the phases of the individual EMFFs, as shown in Fig.\ref{Fig:gmean}. 
As is known, there can be an overall phase that has no effects on the physical observables. Also, one has $G_E^{\Lambda_c}=G_M^{\Lambda_c}$ at threshold. Therefore, we set all the phases to be zero at the threshold.  
There are similarities and differences between the phases of $G_E^{\Lambda_c}$ and $G_M^{\Lambda_c}$. 
For instance, both of them decrease as energy increases. Also, both of them fluctuate. However, the former has more fluctuations than the latter. This is reflected by the relative phase of $G_E^{\Lambda_c}/G_M^{\Lambda_c}$, too. 
The NLO results with cutoffs $\Lambda$=450, 500, 550, 600, 650 MeV are shown in the right column. This time, the difference between different cutoffs looks more apparent than that of the modulus. This urges more measurements from the experiment side to determine the individual EMFFs.

\begin{figure}[htp]
	\centering
	\includegraphics[width=0.99\linewidth, height=0.5\textheight]{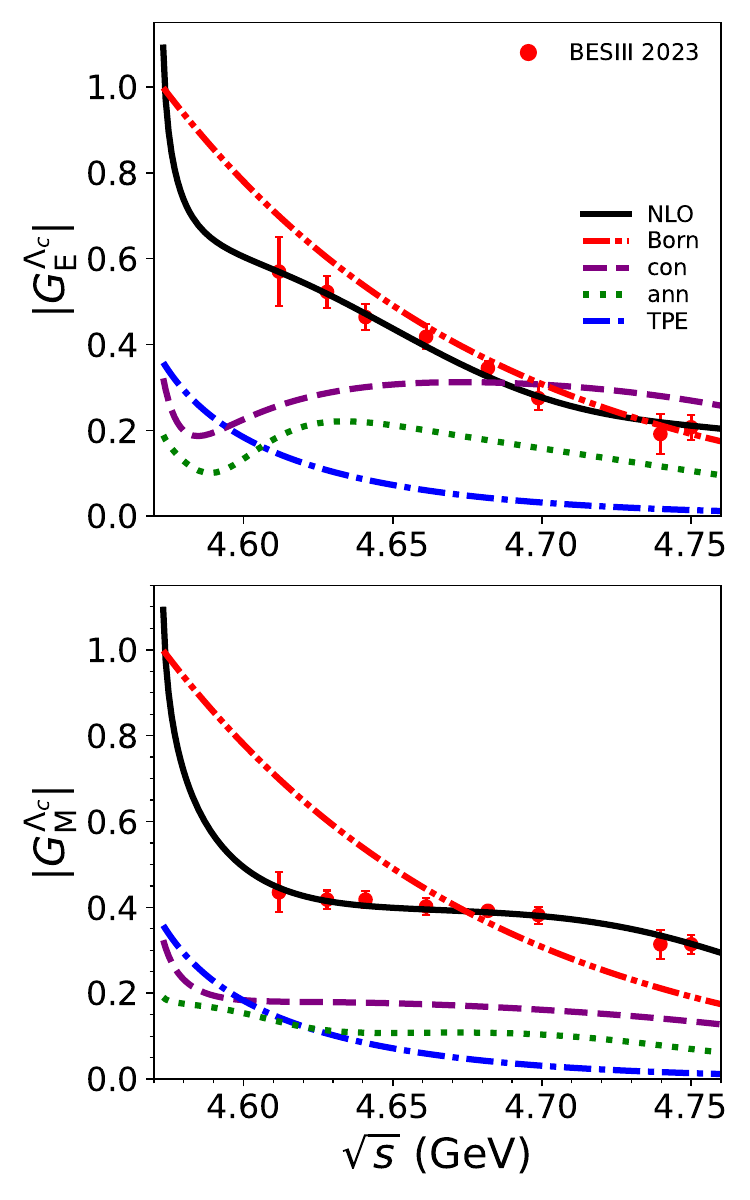}
	\caption{Separated contributions to the individual EMFFs of the $\Lambda_c$. The data is taken from BESIII \cite{Ablikim2018,Ablikim2023} collaboration. The cutoff is chosen as $\Lambda=500$~MeV. }
	\label{Fig:componentGEGM}
\end{figure}
We also give the separated contributions to the modulus of the individual EMFFs, $|G_E^{\Lambda_c}|$ and $|G_M^{\Lambda_c}|$ from the Born term, contact, annihilation, and TPE potentials. 
The one for the electric form factor is at the top, and the magnetic one is at the bottom. 
Both graphs are similar to that of the effective EMFF. 
That is, the Born term dominates the contributions, and the TPE potential has a significant contribution but can not supply the enhancement around the threshold. Instead, the contact potential should be rather crucial in the enhancement for both $G_E^{\Lambda_c}$ and $G_M^{\Lambda_c}$ around the threshold, and the contribution from the annihilation potential can not be ignored for $G_E^{\Lambda_c}$ around the threshold either. Moreover, there are more fluctuations in the $G_E^{\Lambda_c}$ than in the $G_M^{\Lambda_c}$, which are mainly caused by the annihilation part, and the contributions from the contract potential can not be ignored. The former affects the whole energy region, and the latter affects the low-energy region. 
The Born term and TPE are smooth and contribute less to the fluctuation of the individual EMFFs.

\section{Summary}
\label{Sec:IV}
In this paper, we studied the timelike individual EMFFs of the $\Lambda_c$ from the process, $e^+e^-\to \Lambda_c\bar{\Lambda}_c$. The final state interactions of $\Lambda_c^+\bar{\Lambda}_c^-$ re-scatterings are taken into account. The hadronic scattering amplitude is solved by LS equation with the input potentials derived within $SU(3)$ ChEFT. With it, 
the electron-positron annihilation amplitude is constructed according to the DWBA method. The cross-section, modulus of the individual EMFFs, and their ratio up to 4.75 GeV are well described, and high-quality solutions are obtained.
Our amplitudes capture the significant enhancement of cross-section near the $\Lambda_c^+\bar{\Lambda}_c^-$ threshold. It reveals that the contact potential is crucial in the enhancement around the threshold.   
Predictions are made for the individual EMFFs, $G_E^{\Lambda_c}$ and $G_M^{\Lambda_c}$ of the $\Lambda_c$, as well as their ratio $G_E^{\Lambda_c}/G_M^{\Lambda_c}$, including both modulus and phases. Interestingly, we find that the $G_E^{\Lambda_c}$ contains more fluctuations than $G_M^{\Lambda_c}$, which are caused mainly by the annihilation potential in the whole energy region, and also by contact potential in the low energy region. The EMFFs will further study both experimental and theoretical works in the future.  

\section{Acknowledgements}
\label{Sec:V}
This work is supported by the National Natural Science Foundation of China (NSFC) with Grants No.12322502, 12335002, 11805059, 11675051, Joint Large Scale Scientific Facility Funds of the NSFC and Chinese Academy of Sciences (CAS) under Contract No.U1932110, and Fundamental Research Funds for the central universities.

\bibliographystyle{unsrt}
\bibliography{ref.bib}

\end{document}